\documentstyle[aps,twocolumn]{revtex}

\begin{document} 
\input psfig.sty  

\title{A model for mutation in bacterial populations}

\author{R. Donangelo$^1$ and H. Fort$^2$} 
\address{
$^1$ Instituto de F\'{\i}sica,
Universidade Federal do Rio de Janeiro,
C.P. 68528, 21945-970 Rio de Janeiro, Brazil. E-mail:donangel@if.ufrj.br \\
$^2$ Instituto de F\'{\i}sica, Facultad de Ciencias, Universidad de la República,
Igu\'a 4225, 11400 Montevideo, Uruguay. E-mail: hugo@fisica.edu.uy}

\maketitle

\begin{abstract}

We describe the evolution of $E.coli$ populations through a
Bak-Sneppen type model which incorporates random mutations. 
We show that, for a value of the mutation level which coincides
with the one estimated from experiments, this model reproduces
the measures of mean fitness relative to that of a common ancestor,
performed for over 10,000 bacterial generations.

\end{abstract}

\vspace{3mm}


\vspace{3mm}

The last decade has seen a renewed interest in the study of biological
evolution. Besides the painstaking work of analyzing fossil records, 
spanning 10$^{8-9}$ years, there are now experiments performed by
Lenski and co-workers with $E.coli$, which already comprise tens of
thousands of generations~\cite{lt94}.
This has opened an entire ``experimental evolution" area, and their data
are extremely useful to study the long-term evolutionary dynamics of  
populations.

We first briefly review the essentials of Lenski's experiment.
It considered 12 initially identical populations, each of them founded by
a single cell from an asexual clone, propagating in similar environments
during 1500 days, in the following manner. At the beginning of each 24 hour
period, an initial batch of around $5\times10^6$ bacteria is placed in a
growth medium, and, 24 hours later, when the population has increased by a
factor of about 100, which implies $\log _2 100 \approx 6.6$ generations,
the process is repeated by starting a new batch with 1\% of the bacteria
present.

The mean cell volume and mean fitness of bacterial populations relative
to the ancestor $\langle RF \rangle$ were measured every 100 generations. 
The $\langle RF \rangle$ of these populations, in its $g-$th generation
is measured by placing a sample of each of them in contact with its original
ancestor (unfreezing a sample taken at time 0, the first generation), and
measuring the ratio of their rates of increase.
In all the experiments the $\langle RF \rangle$ shows a rapid increase
for $\approx 2000$ generations after its introduction into the experimental 
environment, and then becomes practically constant. 
The average asymptotic value of the relative fitness is
$\bar{\langle RF \rangle} _\infty \simeq 1.48$ \cite{lt94} (the bar 
denotes the average over the 12 populations).  
This behavior may be parameterized by an hyperbolic fit, $f= (A+Bg)/(C+Dg)$,
where $f$ represents the relative fitness of the $g$-th generation and
$A,B,C,D$ are constants.

On the theoretical side, several approaches to the evolutions of
species in interactive systems have been proposed: the well known models
of Kauffman and collaborators for co-evolving species operating at the 
edge of criticality~\cite{kj91}, models inspired on them~\cite{fl92},
\cite{bfl92}, and the Bak-Sneppen (BS) model~\cite{bs93}, among others.

An essential ingredient of evolution theory, complementing the natural
selection mechanism, is the existence of spontaneous mutations which
produce hereditary differences among organisms. Such an ingredient is
not explicitly considered in the standard BS model, but it is clearly
an essential mechanism in the evolution of bacterial cultures.

In this work we modify the BS model in order to include random
mutations so as to explain the $E.coli$ results. 
The reasons for constructing such a model are twofold. 
First, it was experimentally found that the 12 $E.coli$ replicate
populations approached distinct fitness peaks (different asymptotic 
$\langle RF \rangle$ into a band from $\approx 1.4$ to $\approx 1.6$) 
\cite{le2000}, supporting the multi-peaked {\it fitness landscape} of the 
kind assumed in BS-type models. Second, as the initial populations were
identical, and the environment for bacterial growth was kept constant, 
the evolution of these quantities resulted solely from spontaneous
random genetic changes and competition between the different cell
varieties resulting from those changes.

The model thus assumes two kinds of changes, one arising from the
disappearance of the less fit strains and another, completely random, 
that may be attributed {it e.g.} to errors in the replication mechanism. 
All these changes are associated, in the model, to real mutations in the
genome. In the case of the fitness driven changes, such mutations appear
as a two-step process: the extinction of a bacterial strain, and its 
substitution by another, as in the original BS model.
The new random mutations are associated to changes in the genoma unrelated 
to any selection process.

Some clarifying remarks concerning the proper interpretation of the model
in the context of $E.coli$ experiment are necessary:

\noindent {\it i)} Despite the fact that the original BS model considered 
evolution in a coarse grained sense (an entire species was represented by
a single fitness parameter), here we will describe a system of evolving
bacterial populations rather than whole species.

\noindent {\it ii)} The focus of the BS model was to study the dynamic
equilibrium properties of an ecology, {\it i.e.} its Self-Organized
Critical (SOC) behavior after the initial transient.
The data of ~\cite{lt94} consider explicitly the transient evolution
starting from the first bacterial generations. In particular, changes
are observed to be larger for the first 2000 generations, and then
gradually taper off.
Therefore, we consider the evolution of the system from its very initial
state, and not after it has equilibrated.

{\it iii)} In our model we consider a number of cellular automata $N$, which, 
for practical purposes, must take values much smaller than the number of
$E.coli$ in Lenski's experiment ($5\times10^6 - 5\times10^8$).
Below we show that the model has scale invariance properties that
justify this assumption.

As in the case of BS, the model consists of a one-dimensional cellular
automata of size $N$, with cells labelled by a subindex 
$i=1,...,N$. 
Therefore each of these $N$ cells represents a group of bacteria, and
not single individuals. In other words, the system in the model is a
coarse-grained representation of the cell population.
Each vector cell is characterized by a real number between 0 and 1, $B_i$. 
This parameter may be interpreted as measuring the fitness of the ``species",
{\it i.e.} a barrier against changes. 

In order to emulate the experimental condition that each of the 12 
populations was initiated by a single cell from an asexual clone, we
start with the same barrier for all cells, $B_i=0.5$.
The dynamics of the model consists in performing the some operations
at steps corresponding to the time needed for an average E.coli to
divide, {\it i. e.} a generation. Those operations are the following:
at each step, corresponding to the doubling time of the bacteria:
\noindent a) eliminate the cell with the lowest fitness
\noindent b) eliminate, on the average, $Q$ other cells 
\noindent c) replace the barriers associated to the cells eliminated by 
random numbers generated with a uniform distribution in the interval [0,1],
as in the BS model

Results do not depend on the choice of distribution employed in operation c). 
Operations a) and b) are associated to either fitness driven or random
mutations, respectively. 

We should remark that in order to simulate the experiment, one would need
to double the total number of cells at the end of each step, assigning copies
of the barriers associated to the existing cells to the ones created.
Furthermore, every $G\approx 6-7$ steps, the population should be reduced
to its original value $N$ by selecting at random this number of cells among
the total population. 
However, since the mutation probability is found to be constant and
independent of the size of the population, scaling properties \cite{jensen}
allow us to avoid these population doublings and reductions,
keeping the number of cells constant. 
Because of this simplification, the model becomes formally equivalent
to the mean field version of the BS model (MFBS)\cite{fbs93}
However, here the $Q$ barrier changes in c) are interpreted as random
mutations, and not as changes to neighbors of the least fit specimen
in the population as in the MFBS model. 

Since, as in the BS model, after a transient, the model fitness barrier 
distribution self-organizes in a step at $B_c=1/(1+Q)$, its asymptotic mean 
fitness is $(B_c+1)/2=(Q+2)/(Q+1)$, while the mean fitness of the original 
uniform barrier distribution is $0.5$. Therefore, in this model the asymptotic 
relative fitness to the ancestor is 
\begin{equation}
\langle RF \rangle _\infty = \frac {Q+2}{Q+1} \,.$$
\label{eq:RFI}
\end{equation}

We select the value of $Q$ so as to adjust
$\bar{\langle RF \rangle} _\infty \simeq 1.48$.  
Hence, from (\ref{eq:RFI}) we obtain $Q \simeq 1.1$.
It is interesting to note that the model suggests an approximately equal
number of fitness driven and random mutations for the E.coli under the
conditions of Lenski's experiment.

While $Q$ determines the asymptotic value of $\langle RF \rangle _\infty$,
the number of cellular automata, $N$, is determined by the empirically
observed mutation rate per replication $\mu$, which was estimated as 
$\mu \approx 0.002$ per replication~\cite{d91}, \cite{sc96}.
Since in the model we have on the average $Q+1=2.1$ changes per generation,
then the number of cells in the simulation should be $N\approx 1000$.
In this way, the two parameters of the model, $Q$ and $N$, are fixed so
as to reproduce the experimentally observed asymptotic fitness and
mutation rate found in the experiment.
The simulations were performed for $Q=1.1$ and for 
several values of $N$ in the interval $500 \leq N \leq 2000$.

The agreement between the model and experiment is quite reasonable.   
In Fig.~1 we plot the $\langle RF \rangle$ trajectory every 500 
generations for $N=1000$ ($\Box$)
and $N=1500$ ($\triangle$) and 3 of the Lenski {\it et al.}
best hyperbolic fits to 3 (of the 12) sets of data for the $E.coli$ 
experiments.

\begin{center}
\begin{figure}[h]
\centering
\psfig{figure=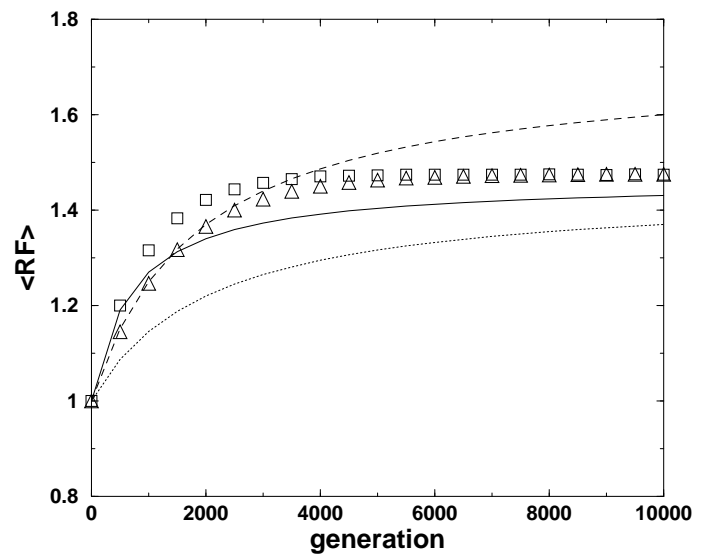,height=7.5cm}
\caption{Trajectories for mean fitness relative to the original ancestor during
10,000 generations.
Averages from 1000 numerical simulations for $Q=1.1$ for $N=1000$
(squares) and $N=1500$ (triangles) compared with the 3 best hyperbolic
fits (lines) to data of 3 of the 12 corresponding experiments.}
\label{fig1}
\end{figure}
\end{center}

In order to analyze the initial rapid grow of $\langle RF \rangle$, 
in Fig.~2 we plot the experimental data and the model results for 
the first 2000 generations, every 100 generations. 
Once again notice the overall agreement of the model with 
experimental E.coli data. 

\begin{center}
\begin{figure}[h]
\centering
\psfig{figure=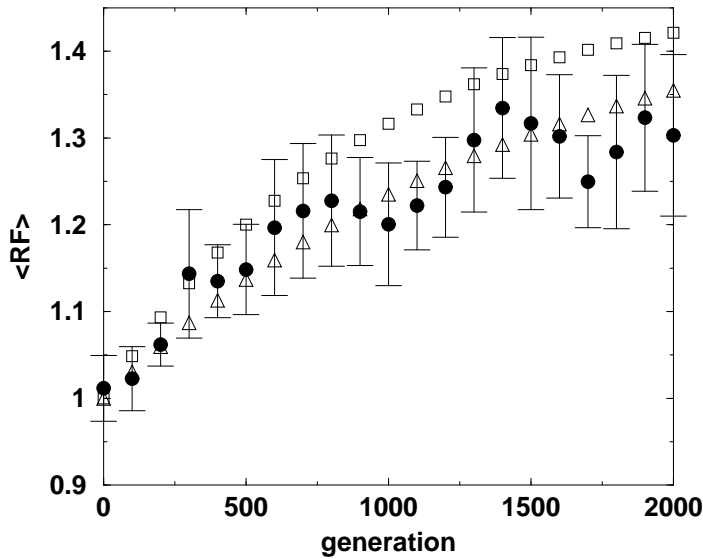,height=7.5cm}
\caption{Finer scale analysis of the trajectories of the mean
fitness relative to the ancestor for the 2000 initial generations.
Experimental data (filled circles), and model averages from 1000 numerical
simulations with $Q=1.1$: $N=1000$ (squares) $N=1500$ (triangles).
The standard deviations of the experimentally measured relative 
fitness are indicated as error bars in the data points. 
The model calculations have negligible dispersion in the scale
of the plots presented in this work.}
\label{fig2}
\end{figure}
\end{center}

It was suggested in~\cite{lt94} that 
periods of stasis, characteristic of punctuated evolution, might 
be present in the data. 
Although the relatively large experimental error bars could make
the data consistent with a monotonic increase like the one
predicted by the model, we believe that further studies are 
needed to settle this issue. 

The results presented thus far assume that there is no neighbor relation
between different strains of E.coli in the system. However, as in other
ecological system interdependencies among species arise, one should
explore the possibility that also in this case they exist.
To do this, we think it is illustrative to present here also the
results of a variation of the standard BS-type model.
In this variation, as in the original model, at each time step the
changes occur at the cell with lowest barrier and its two neighbors.
In addition to this fitness driven form of evolution we include, 
with probability $p$ per time step, a similar change in a randomly 
chosen cell. In this way the number of changes per time step is 
$Q=2+p$ (the two neighbors of the cell with minimum barrier plus,
with probability $p$, a cell at a random location). We denote this 
version of the model as BS$+p$.

The stationary properties of this BS$+p$ model will be presented
elsewhere \cite{df01}. We have observed that the barrier distribution
self-organizes into a step function at a position $B_c$ which decreases 
as the parameter $p$ increases from $B_c\simeq0.667$ for $p=0$ (the
standard BS model) to $B_c\simeq 0.22$ for $p=1$.
Thus $\langle RF \rangle _\infty$ lies in the interval between 1.22
(for $p=1$) and 1.667 (for $p=0$). 
We found that $\bar {\langle RF \rangle }_\infty \simeq 1.48$ may 
be adjusted taking $p\simeq 0.2$.
Hence, here the purely random mutations are taken with a weight
proportional to 0.2, while those mutations related to natural selection
are proportional to 3.

In this BS$+p$ version, in an analogous way as it happens in the MFBS version, 
while $p$ determines the asymptotic value of $\langle RF \rangle _\infty$,
the number of cellular automata, $N$, is determined to fit with the
estimated $\mu \approx 0.002$.
Since, we now have, on the average, 3.2 changes per time step,
in order to get roughly the same mutation rates in the simulation and in
the experiment we should take $N \simeq 1500$.
In Fig.~3 we plot the $\langle RF \rangle$ trajectory every 500 
generations for $N=1000$ ($\Box$)
and $N=1500$ ($\triangle$) and 3 different hyperbolic fits of \cite{lt94}.
Notice the good agreement with the Lenski {\it et al.} hyperbolic fit to
data corresponding to the ``A-1" experiment \cite{le2000}. 

\begin{center}
\begin{figure}[h]
\centering
\psfig{figure=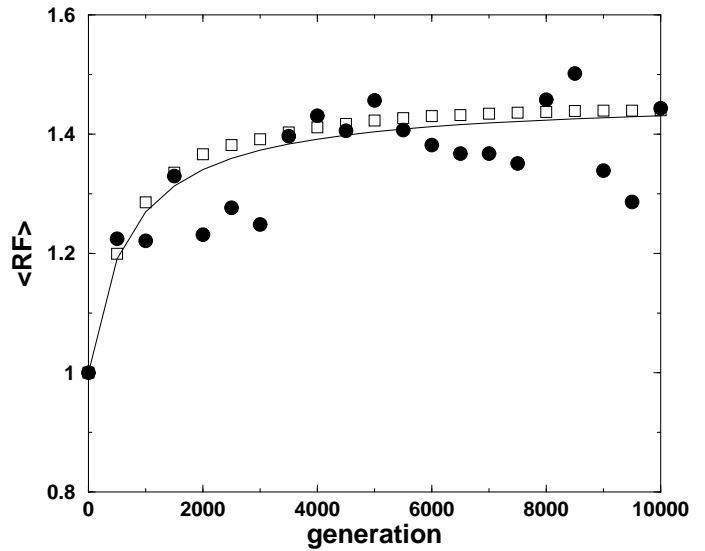,height=7.5cm}
\caption{Trajectories for mean fitness relative to the original ancestor during
10,000 generations, in the BS$+p$ model.
Averages from 1000 numerical simulations for $p=0.2$ for $N=1500$
(squares) compared with the data (filled circles) and their best hyperbolic
fit (line). See text for further details.}
\label{fig3}
\end{figure}
\end{center}

In the case of the E.coli, the MFBS model appears, in principle, more 
reasonable, as it does not seem plausible to have interdependencies among 
closely related bacterial strains. The BS$+p$ model could be applicable to
other situations, where different microorganisms coexist and are interdependent. 
Our point here was to show that it is not possible to distinguish between the
two models based solely on the measurements of the evolution of the fitness. 

To conclude, when considered during the transient from the initial ordered
distribution, BS models with random mutations were shown to qualitatively
reproduce the experimental results of Lenski and co-workers. 
The inclusion of random mutations, besides making more realistic the
models, is required to get quantitative agreement with the experimental 
results, both for the transient and the asymptotic regime.
While both fitness driven and random mutations were shown to be needed, 
their relative importance remains an open question.
One should remark that the calculations presented here are just an
starting point in the exploration of this complex biological system.
In particular, the existence of stasis regions, suggested by the data, 
remains another open issue.

\vspace{2mm}

Work supported in part by PEDECIBA (Uruguay) and CNPq (Brazil).
We thank L. Acerenza, G. Gonz\'alez-Sprinberg and K. Sneppen 
for valuable discussions and suggestions.

\end{document}